\documentclass{article}

\usepackage{tikz,amsmath,amssymb}
\usetikzlibrary{shapes,backgrounds}
\usepackage{verbatim}
\usepackage{algorithm}
\usepackage[noend]{algpseudocode}
   \usepackage{multirow}      
    \usepackage{multicol} 
\usepackage[font=scriptsize]{subcaption}
\algnewcommand{\LineComment}[1]{\State \(\triangleright\) #1}

\begin{document}
	
	\title{CUPSMAN: Control User Plane Separation\\Based Routing in Ad-hoc Networks}
	
	\author{Doganalp Ergenc, Ertan Onur
		\\Department of Computer Engineering,\\Middle East Technical University,\\06800 Ankara Turkey \\ \{ergenc.doganalp, eronur\}@metu.edu.tr}

	\maketitle
	
	\begin{abstract}
			Separation of user (data) plane from the control plane in networks helps scale resources independently, increase the quality of service and facilitate autonomy by employing software-defined networking techniques. Clustering introduces hierarchy in ad hoc networks where control functions can be carried out by some designated cluster heads. It is also an effective solution to handle challenges due to lack of centralized controllers and infrastructure in ad-hoc networks. Clustered network topologies gain a significant amount of scalability and reliability in comparison to flat topologies. Different roles that nodes have in a clustered network can be effectively used for routing as well. In this paper, we propose a novel plane-separated routing algorithm, Cluster-based Hybrid Routing Algorithm (CHRA). In CHRA, we take advantage of the hierarchical clustered structure through control and user plane separation (CUPS) in mobile ad-hoc networks. In the cluster neighborhood with a particular size, a link-state routing is used to minimize delay, control overhead, and also utilize energy consumption. For facilitating the communication with distant nodes, we form a routing backbone that is responsible for both control and data messages. The results show that CHRA outperforms its opponents in terms of fair energy consumption and end-to-end delay.
	\end{abstract}

	\section{Introduction} \label{sec:intro}
	
{S}{calability} in large-scale ad-hoc networks is a crucial requirement for providing a high performance~\cite{khakpour2017using}. Clustering that can generally be defined as grouping of nodes based on some common properties, is commonly employed for achieving scalability in such networks~\cite{cooper2017comparative}\cite{rossi2017stable}. Nodes in a clustered topology are categorized into different roles according to their functionality and those roles can be effectively used for routing as well. Cluster-based Hybrid Routing Algorithm (CHRA) is a novel routing algorithm that takes advantage of the clustered structure of a network, basically using the nodes which are cluster heads (CH) and gateways to manage routing and separate control and user planes. 

{Control plane} is defined as a set of nodes and procedures that are responsible for routing decisions using control messages. In contrast, user or {data plane} includes the nodes which are forwarding data packets (instead of control packets used in the control plane) in an end-to-end communication via any pre-determined path or route. In CHRA, we define the term \textit{cluster sight area} (CSA) that covers maximum \textit{n}-CH-hop distance where \textit{n} is a design parameter to limit the area. In this area, we proactively establish routes and minimize routing control messages. For the end-to-end communication outside a CSA, we used an on-demand approach to find a route between source and destination through the routing \textit{backbone} which consists of neighbor cluster heads and gateways.

There is a number of routing algorithms which are designed for the clustered networks. Such algorithms usually fall under the hierarchical routing category. Cluster-based Routing Protocol (CBRP) \cite{ietfcbrp} is one of the first hierarchical algorithms. In CBRP, CHs send route requests to their neighbor clusters via gateways when a member of the related cluster needs to communicate with a remote one. Intra-cluster communication is based on broadcasting instead. Most of the other routing protocols take a similar approach with CBRP. Cross-CBRP \cite{Jahanbakhsh2008} evaluates the ratio between power levels of two successive signal for clustering phase to elect more reliable cluster heads and then routing is performed through them as in CBRP. CLACR \cite{Wang2009} and CMDSR \cite{An2005} take advantage of some central infrastructures (managers and servers) to discover topology and find more accurate routes. However, their applicability is limited to some specific scenarios. In contrast, the topology discovery is performed in a homogeneous network using a BGP-like domain division approach in CIDR \cite{Zhou2009} and cross-domain or inter-cluster communication is done via cluster heads and gateways like CBRP. As an alternative approach, ZHLS \cite{Lu1999} uses GPS for clustering to group closer nodes and routing to detect further nodes and obtain related routes.
	 
	There are also hybrid routing algorithms that effectively use hierarchy in clustering structure. HCR \cite{Niu2006}, for instance, uses proactive approach for intra-cluster communication and continuously updates routes. For inter-cluster communication, it behaves like all other algorithms presented here and uses CHs and gateways directly.

Details of the clustering algorithms are discussed in different studies before \cite{cooper2017comparative}\cite{rossi2017stable} \cite{Vodopivec2012}. Independent from those technical details (e.g how CHs are selected or clusters are form and maintained), CHs in hierarchical structures naturally bring a variety of advantages for management of the network. However, using them to find routes and forward data together exhausts those specially-selected nodes, i.e cluster heads and gateways. Moreover, many possible routes that can be defined by ordinary nodes are neglected while focusing on cluster heads. In this paper, our main contributions are formation and maintenance of separate planes; control and data plane in mobile ad-hoc networks without any special node (e.g. having longer transmission ranges, GPS) or central predeployed mechanism (e.g. routing servers and managers). We aim to separate control and data planes, i.e routing and forwarding for (1) a fair energy consumption and (2) traffic-load distribution between nodes, and (3) a low end-to-end delay using all other nodes for forwarding as well. 

	The rest of the paper organized as follows: In Section \ref{sec:CHRA}, working principles of CHRA are detailedly shown. In Section \ref{sec:results}, simulation environment and performance metrics are presented, and the performance of CHRA is discussed comparing with its opponents, backbone routing and AODV. Finally, Section \ref{sect:conclusion} represents the conclusion and future works to extend this study.

\section{Cluster-based Hybrid Routing Algorithm}\label{sec:CHRA}

In this section, the key terms and working principles of CHRA are  explained considering different distances between source and destination nodes. Communication in short distances i.e both source and destination nodes are located in same CSA, and long distances which are more than \textit{n}-CH-hop require a different kind of routing approaches to ensure low communication delay and overhead with a high packet delivery ratio. 

\subsection{Terminology} \label{sect:terms}
The backbone and cluster sight area are the core definitions in CHRA. They construct the control plane of the network and take roles in both clustering and routing. Besides, note that the definition of the backbone is conceptually similar to structures in other cluster-based routing protocols which use CHs and gateways to maintain both control and data traffic as presented in Section \ref{sec:intro}. In this section, these terms are defined for the comprehensibility of the algorithm.

\subsubsection{Backbone}

\begin{figure}[htb!]
		\centering
		\includegraphics[scale=0.4]{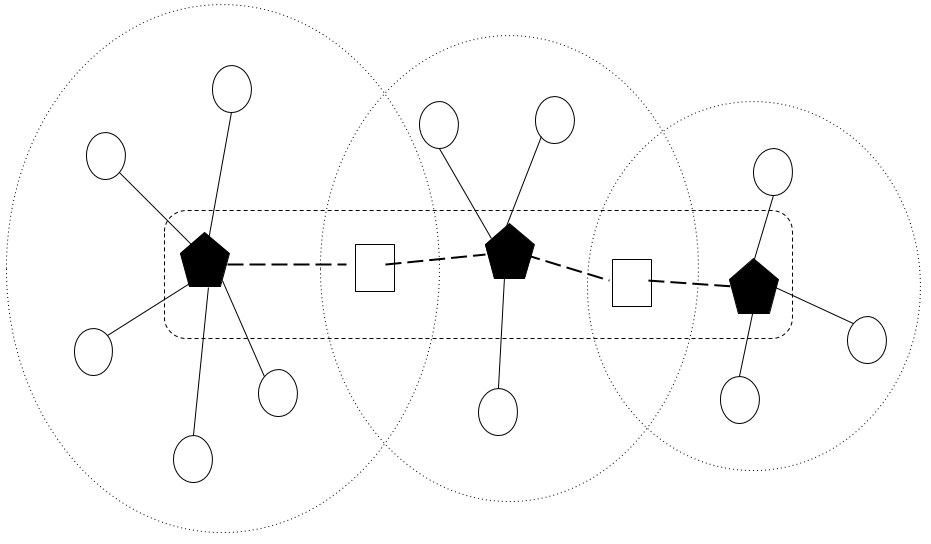}
		\caption{The backbone is constructed by cluster heads and gateways.}
		\label{fig:backbone}
\end{figure}

Instead of flooding through the whole network, routing control messages are being forwarded via a specific set of nodes which are CHs and gateways as shown in Fig.~\ref{fig:backbone}. Except isolated clusters which have no nodes connected to any node from a different cluster, all CHs are connected to the others through gateways. Therefore, in a fully connected network -in terms of clusters-, CHs and gateways form a continuous structure, which is called the backbone. Since CHs have topology information of their own clusters, they can easily make decisions for routing which is destined to any cluster-member node. Therefore, maintenance and discovery of the routes are narrowed down to the backbone, which also forms the control plane. From this perspective, control plane and backbone terms can be used interchangeably for both routing and clustering processes.

\subsubsection{Cluster Sight Area} \label{sect:sightarea}

\begin{figure}[htb!]
		\centering
		\includegraphics[scale=0.4]{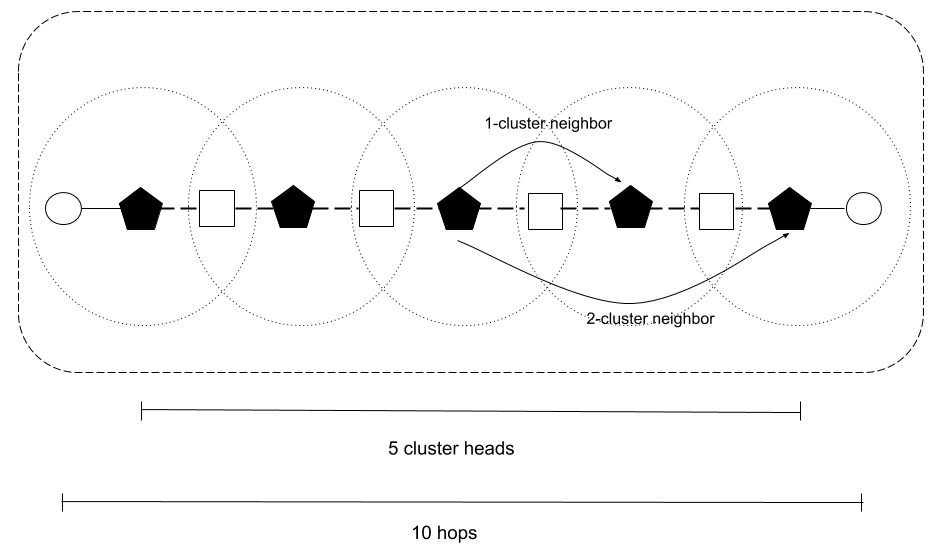}
		\caption{Cluster sight area covers maximum 10-hop.}
		\label{fig:sightarea}
\end{figure}

Apart from the neighborhood of nodes in a flat topology, it is considered for the neighborhood of cluster heads in a hierarchical network. In Fig. \ref{fig:sightarea}, diamonds, squares and circles represent CHs, gateways and ordinary nodes respectively. Taking the CH in the middle as the pivot, each CH discovers its sight area considering the clusters in -at most- \textbf{\textit{n}-CH-hop} neighborhood as shown in Fig. \ref{fig:sightarea}. A CH-hop represents the distance between two clusters in terms of cluster heads, that is, if two cluster heads can communicate via a single gateway, their clusters are direct neighbors in a 1-CH-hop distance. For instance, Fig. \ref{fig:sightarea} represents a 2-CH-hop neighborhood where the radius of CSA is 2-CH-hop. In this limited area, all topology is known by all CHs, i.e any link between nodes is identified by each cluster head. Naturally, each cluster head discovers and maintains its own cluster's topology regularly with periodic clustering control messages. Similarly, each CH sends this local topology information with inter-cluster sight area messages (SAM) to its neighbor CHs via gateways. To reduce control overhead for maintenance of the area, inter-cluster control messages are sent in different periods with a fish-eye approach. That is, while the control messages are sent in every $T_{SAM}$ seconds to 1-CH-hop neighbors, they are sent to n-CH-hop neighbors in $nT_{SAM}$ seconds. Eventually, each CH has more fresh and reliable topology information about closer clusters. In this manner, CSAs are maintained proactively by CHs, as a natural extension of clusters. $T_{SAM}$ is chosen as 3 seconds for simulations and the effects of its frequency are briefly discussed in Section \ref{sec:results}.

For simulations, we selected \textit{n} for CSA as 2. Note that, 2-CH-hop distance is not mandatory but a design issue. The most primal fish-eye approach for inter-cluster information exchange contains (at least) 2-CH-hop distance so that a pivot cluster head is able to discover neighbor topologies with changing frequencies which are proportional to distance. The upper bound for such structure is the whole network, i.e sending topology information to all other clusters. In contrast, 2-CH-hop constructs the minimal structure and eventually minimum control overhead for topology discovery. The implicit relationship between CH-hop and regular node neighborhood is also simple, assuming that cluster heads are connected to each other via gateways (not directly connected), \textit{n}-CH-hop contains $(4n + 2)$-hop paths at most.

The main reason for the construction of CSA is creating sense of a smaller network which is relatively easy to maintain. Since proactive maintenance of the network-wide routes is costly, a full-discovery only in a smaller area decreases the delay in end-to-end communication and utilizes control overhead for routing. Therefore, CSA is an effective yet easy-to-maintain structure depending on its size.

\subsection{Algorithm} 

In CHRA, two major distance-dependent approaches are taken in routing process. In-area communication represents the short distances inside a CSA. In contrast, long-distance communication means end-to-end communication outside the CSA.

\begin{table}[ht!]
\centering
\caption{Different approaches in CHRA.}
\label{tab:wrapup}
\begin{tabular}{|c|c|c|c|}
\hline
                & Plane Separation & Routing Type              & Routing Main.  \\ \hline
In-area Com.    & Yes              & Link-state & Semi-proactive \\ \hline
Long-dist. Com. & No               & Distance Vector           & Reactive       \\ \hline
\end{tabular}

\end{table}

The hybrid approach in CHRA brings different techniques taking advantage of the clustered structure. Table \ref{tab:wrapup} summarizes those techniques. Only in in-area communication, data plane is used for the data transfer and the routes are found with routing control messages through the backbone. Therefore, effective use of plane separation is observed there. Since the complete topology is discovered in a small area i.e CSA, the routes including end-to-end paths (EEP) from source to destination can be discovered. Even if CSAs are proactively maintained, routes are still drawn on-demand; therefore it is regarded as a semi-proactive technique. In contrast, long-distance communication is totally up to backbone routing and data carriage on-demand. Thus, its routing type is reactive.  In this section, both approaches are explained with examples. Algorithm \ref{alg:routeDiscovery} briefly shows hybrid routing process as well. 

\begin{algorithm}
    \caption{Hybrid route discovery process.}
    \label{alg:routeDiscovery}
    \begin{algorithmic}[0] 
        \Procedure{RouteDiscovery}{}
	\State Send an RREQ to CH containing destination address      
    \LineComment{CH checks;} 
	\If {Destination node in CSA}
		\State Send RREP containing the shortest EEP to previous-hop
	\ElsIf {Destination node in distance vector}
		\State Send RREP containing next-hop information to previous-hop
	\Else
		\State Forward RREQ to other CHs via gateways		
	\EndIf      
	
	\LineComment{Source node checks;} 
	\If {No RREP received until timeout}
		\State Start route discovery again
	\Else
	\If {RREP contains full EEP}
		\State Update routing table to keep the shortest EEP
	\ElsIf {RREP contains only next-hop}
		\State Update distance vector
	\EndIf  
	\EndIf    	
   
        \EndProcedure
    \end{algorithmic}
\end{algorithm}

\subsubsection{In-area Communication} 

\begin{figure}[htb!]
		\centering
		\includegraphics[scale=0.4]{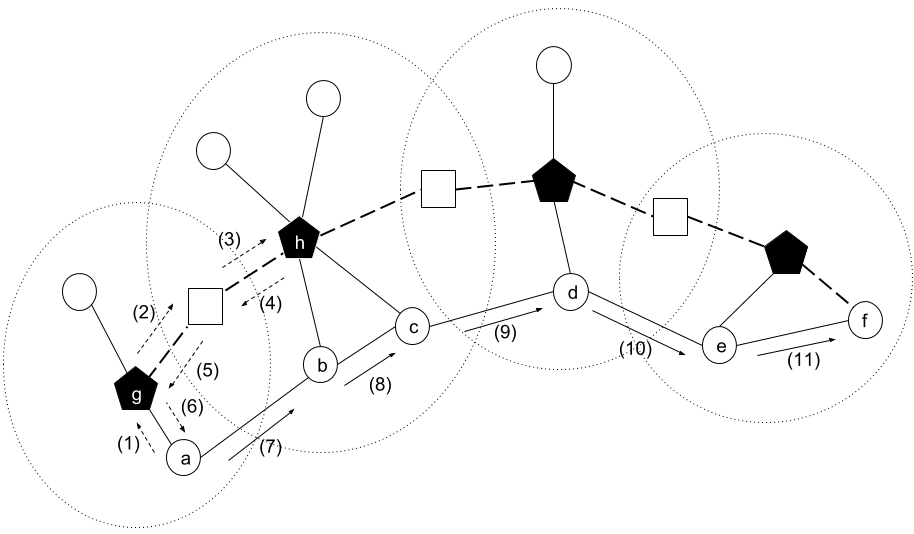}
		\caption{Routing inside a CSA.}
		\label{fig:shortcom2}
\end{figure}

In a CSA, since at least one CH knows the complete topology of this area, it is possible to find an end-to-end path (EEP) other than the backbone's itself. In this case, control and data planes are separately considered for finding a route and forwarding data respectively. In this section, route discovery  and error handling methods in short-distance communication are explained.\\ \\
\textbf{Route Discovery:} In Fig. \ref{fig:shortcom2}, an in-area communication is illustrated with no-use of the control plane (i.e backbone) for data forwarding. In the figure, node \textit{a} and \textit{f} are source and destination nodes respectively. To find the route going to \textit{f}, \textit{a} sends a route request (RREQ) to its CH \textit{g} with packet (1) containing connection demand to \textit{f}. Each CH stores a \textbf{visibility matrix} that contains adjacent nodes and represents the complete neighborhood in the CSA. Note that, it is proactively formed using the topology information in periodic SAM packets. When the CH receives an RREQ, it checks the visibility matrix if the destination node is visible first, which is \textit{f} in this scenario. If it is visible i.e exists in the matrix, the cluster head runs Dijkstra's shortest path algorithm on the visibility matrix and finds the shortest path independent from the backbone. Since node \textit{f} is not placed in visibility matrix of \textit{g}, RREQ in packet (1) cannot be responded directly. Instead,  \textit{g} forwards the RREQ to neighbor CHs via gateways through the backbone.  Packets (2)-(3) represents forwarding RREQ to the neighbor CH, \textit{h}. It is aware of the whole topology shown in Fig. \ref{fig:shortcom2} since every node is placed in 2-CH-hop range with respect to \textit{h}. Therefore, it can find a complete route from \textit{a} to \textit{f} by running Dijkstra's shortest path algorithm on its visibility matrix. 

Afterwards, it sends back the EEP \textit{a}-\textit{b}-\textit{c}-\textit{d}-\textit{e}-\textit{f} to node \textit{g} with packets (4)-(5), and \textit{g} notifies the source node \textit{a} with a routing response (RREP) containing the demanded route. Note that the process in control plane is finished after route request and response messages have arrived. Finally, node \textit{a} forwards the packets tailing the whole path to \textit{b}, and the forwarding process is continued hop-by-hop through the packets (4)-(7) in data plane which consists of ordinary nodes. If an intermediary node does not store this particular path, it also caches it for future uses and forwards to next node with actual data. Otherwise, it forwards only payload dropping the path information. Note that intermediary nodes use cached routes for only data forwarding. It means that they do not use an indirectly obtained path for initiating an end-to-end communication as a source node. The reason for this restriction is explained in the next part, Route Error.\\ \\
\textbf{Route Error:} When a broken link exists in the backbone, it is relatively easy to detect since CHs have a periodic message exchange scheme for control messages of any clustering algorithm. However, it is not always possible to detect a broken link between two ordinary nodes. In such cases, any route containing the broken link losses its validity. Other nodes using related invalid routes need to be informed such losts using minimum number of control messages. Therefore, control and data plane separation requires a routing error mechanism to continuously manage routes in the data plane. 

\begin{figure}[htb!]
		\centering
		\includegraphics[scale=0.4]{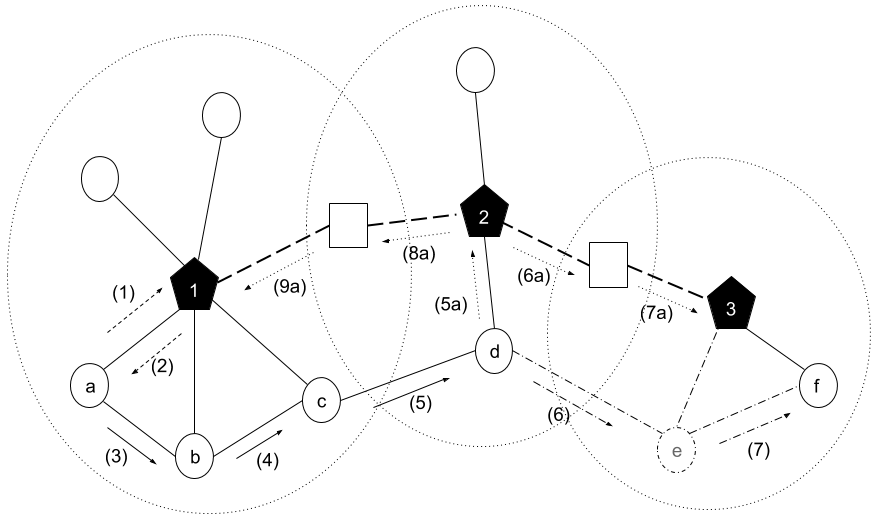}
		\caption{Routing error in data plane.}
		\label{fig:shorterror}
\end{figure}

Fig. \ref{fig:shorterror} shows a routing error scenario in the data plane. In the scenario, node \textit{e} is not there anymore due to mobility, or a node crash. Since nodes periodically send keep-alive messages to maintain a clustered structure as a common nature of clustering algorithms, we assume that its neighborhood becomes aware of the loss in a while depending on the cluster update scheme. 

When node \textit{a} sends data using the source route that is obtained from its CH with the control messages (1)-(2), packets are forwarded through node \textit{d}, and node \textit{d} detects that the EEP is actually broken since node \textit{e} is off. In this scenario, it deletes any recorded source route which node \textit{e} is included and send a route error (RERR) packet to its cluster head with the control message (5a). The first RERR packet (5a) contains the source of the route (node \textit{a}), identifier of the lost node \textit{e} and timestamp for the loss. When a CH receives an RERR packet, firstly it deletes the lost node from its sight area and all recorded routes containing that node from both list of EEPs and distance vector. Then, construct a list of source nodes (\textbf{source notification list}) that requested any of the deleted routes. Note that, destination nodes of those routes also added to the list since they record reverse route as well. Adding the source notification list to the route error message, it forwards this RERR to all neighbor cluster heads. After the first RERR packet, each cluster head applies same procedure with a difference, they also forward RERRs to the nodes which are in source notification list and update this list. Eventually, all cluster heads are informed about a broken link, and also related source nodes do not use these source routes anymore. Note that, this method is only applicable when intermediary nodes are not allowed to use cached route to initiate a connection. If they do so, source nodes for related routes cannot be tracked and RERR messages need to be broadcasted frequently in high-mobility networks.

Note that, while a node is deleting a lost node from its sight area, another CH which is not aware of this loss yet may send a topology update message containing the lost node. Therefore, it is necessary to be able to decide freshness of the information so that CHs can suggest a valid end-to-end route. Each node which deletes an EPP records the lost node in its \textbf{node ban-list}. Each entry in node ban-list contains the ID of a lost node and timestamp when its loss is detected. In case of a topology update, a node firstly checks its ban-list if any node in topology message appears in its ban-list. If any exists, it checks the timestamp in related ban-list record to evaluate how long it has been since the node is lost; if more than $T_{ban}$ seconds passed after lost, then topology update is considered as "fresh" otherwise lost node-related part of the topology update message is discarded. $T_{ban}$ is directly related to mobility level of a network and in our test scenarios, it is determined as 10 seconds.

When RERR messages are propagated through the network via the backbone, it is possible to get same RERR packets for a node. Because, there are multiple paths to access CHs, gateways and source nodes. Assuming there is no isolated cluster, all CHs are connected forming the backbone and it means all of them would receive RERR packet at least once. Each CH records the sequence number of RERRs and directly discards duplicates. Besides, since EEPs are defined in maximum $(4n+2)$-hop (where CSA has a \textit{n}-CH-hop radius), TTL of RERR packets for source nodes is limited to $(4n+2)$. Eventually, the duplicate discard and TTL limitation minimize flooding of the RERR packets.\\ \\
\textbf{Route Repair:} There is also an alternative method to overcome excessive RERR handling in high-mobility networks, that is route repair. When a node detects a broken link, it is able to repair the broken part before sending a RERR packet.

\begin{figure}[htb!]
		\centering
		\includegraphics[scale=0.4]{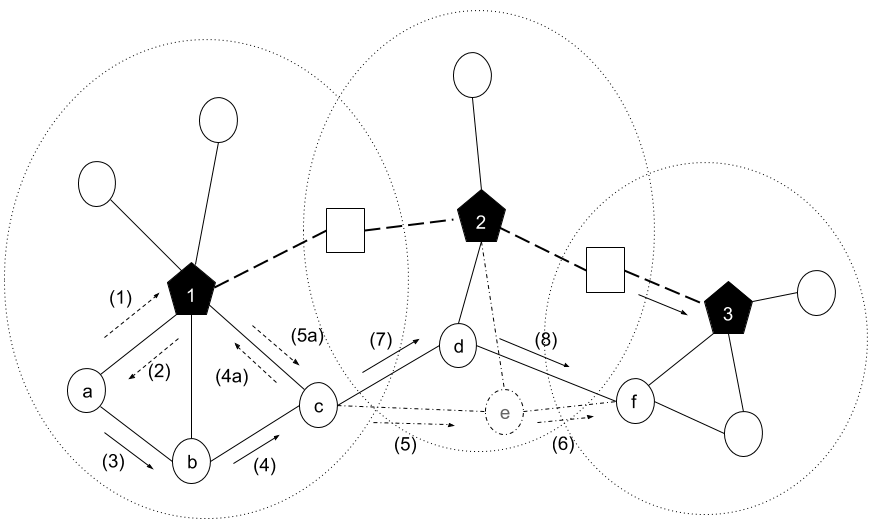}
		\caption{2-hop repair.}
		\label{fig:repair1}
\end{figure}

In CHRA, there are two types of route repair mechanism. The first one is \textbf{2-hop repair} and it aims minimum control overhead and change in an existent route  for repair. The other one is \textbf{full repair} which aims low delay with a more controllable approach. Fig. \ref{fig:repair1} shows an example of 2-hop repair. The main idea behind this type of repair is, instead of finding an alternative route after a broken link or in an end-to-end fashion, only an alternative next-hop node is searched. In this sense, related route is patched with minimum effort and it is not required to spread RERR messages through backbone for a lost intermediary node in the route. In Fig. \ref{fig:repair1}, the EEP which is decided for communication between node \textit{a} and node \textit{f} is \textit{a-b-c-e-f}. When node \textit{c} detects the broken link to \textit{e}, it sends repair request (RPREQ) to its cluster head with (4)-(5). Since the cluster head (1) can observe the whole topology presented in the figure, it directly looks for an alternative path going from \textit{c} to \textit{f}, instead of \textit{a} to \textit{f}. Note that, looking an alternative route from \textit{c} to \textit{f} is not for directly finding a partial path to destination, it is finding another next hop that completes the previous route \textit{a-b-c-e-f}. The only update in the route is forwarding through node \textit{d}, instead of node \textit{e}. Whether node \textit{a} is aware of the loss of node \textit{d} or not, \textit{c} repairs the path without announcing it to the network but its cluster head. During repair, data packets are cached in the latest node (node \textit{c} in this scenarios) before the broken link. Eventually, minimum number of control messages is emitted and it leads both higher resource utilization and low delay communication. However, it is not always possible to perform 2-hop repair considering an alternative next hop. Such cases lead us to full repair method which triggers more changes in the network.

\begin{figure}[htb!]
		\centering
		\includegraphics[scale=0.4]{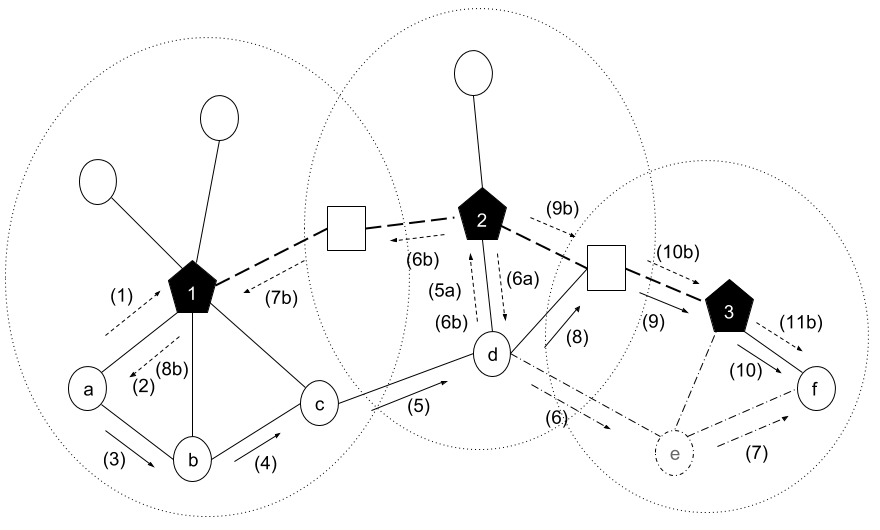}
		\caption{Full repair.}
		\label{fig:repair2}
\end{figure}

Fig. \ref{fig:repair2} shows the alternative method which is used if 2-hop repair cannot be performed. Full repair basically tries to find a full sequence of nodes after broken link connecting to the destination node. It is different than finding a new EEP since it only completes the path after a broken link. Therefore, the EEP is repaired without disconnection but the loss in route is announced to the network as explained above. In the figure, when node \textit{d} detects the broken link, it sends an RPREQ packet to its cluster head with (5a)-(6a). Since CH (2) cannot find a 2-hop repair alternative (a single alternative node instead of node \textit{e}), it draws a totally different path to replaced with the broken part where (8)-(10) forwarded through. Additionally, the cluster head sends RERR packets through the backbone including updated path and route identifier of broken link with (6b)-(11b) so that related nodes (source and destination nodes using that route) can update the older route. Note that, while the invalid route is announced to the network in full repair, it is not a case in 2-hop repair. Because maintenance of a one-hop updated path is relatively easier than a multi-hop path. That is, when full repair is performed, any other repair in formerly-repaired parts leads nodes to maintain multiply-repaired routes without awareness of other nodes including CHs. To avoid side effects in such scenarios and keep the maintenance easier, any changeover in routes is also spread to the rest of the network via the backbone.

Lastly, there could be such scenarios where route repair is not possible at all, however RPREQ packets are sent in any case since repair cannot be performed without CHs which manages the sight areas. Therefore, nodes are waiting for RPREQ response (RPREP) for a limited time, then drops cached data packets if related RPREP is not received. The trigger of RERR packets is performed by the CHs which receives RPREQ, but cannot repair the broken part. Algorithm \ref{alg:routeRepair} briefly concludes the procedure in case of route errors. 

\begin{algorithm}
    \caption{Route repair process for end-to-end paths in data plane.}
    \label{alg:routeRepair}
    \begin{algorithmic}[0] 
        \Procedure{RouteDiscovery}{}
	\State Send a RPREQ to CH containing missing node and invalid path identifier     
    \LineComment{CH checks;} 
	\If {There is an alternative node to patch the route}
		\State Send RPREQ to originator node containing patched route
	\ElsIf {A new path exists from originator node to destination node}
		\State Send RPREQ to originator node containing new path drawn from originator to destination node
		\State Send RERR to other CHs and source nodes via backbone
	\Else
		\State Send RERR through the backbone containing missing node and invalid path identifier
	\EndIf      
	    
        \EndProcedure
    \end{algorithmic}
\end{algorithm}

\subsubsection{Long-distance Communication} \label{sect:longcom}

\begin{figure}[htb!]
		\centering
		\includegraphics[scale=0.4]{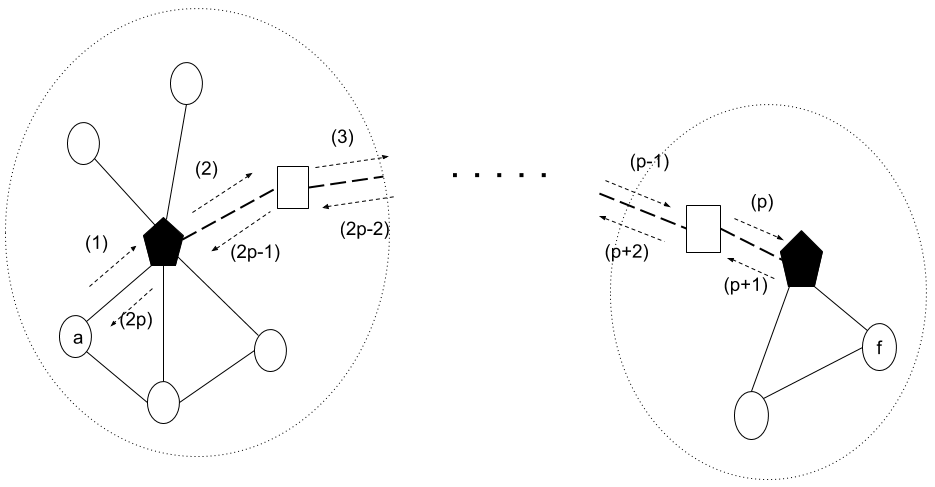}
		\caption{Communication in longer distances is constructed on the backbone.}
		\label{fig:longcom1}
\end{figure}

For communications outside a CSA, there is not a single CH that can find an EEP. Therefore, instead of separating data plane and control plane, data messages are forwarded through the backbone. That is, both control messages and data packets are forwarded via CHs and gateways. In Fig. \ref{fig:longcom1}, source node \textit{a} is lying at a distance $p+1$ from destination node \textit{f} where $p \geq 4n+2$. When node \textit{a} sends an RREQ to its CH, it cannot find a complete route and forwards the RREQ to neighbor clusters via gateways. Since no intermediary CH has both source and destination nodes in its visibility matrice, the RREQ is forwarded it reaches to the cluster where the destination node \textit{f} belongs. Packets (1)-(p) represent forwarding process through the CH of the destination node's cluster. Afterward, since the CH knows all members of its cluster, it sends an RREP back to the node which related RREQ is come from. In each node on the backbone, the node which RREP is sent is recorded so that only next hop for the related route is known, i.e it constructs a distance vector for each different destination. In this sense, this approach is similar to AODV routing which is only constructed onto the backbone. Packets (p+1)-(2p) show the RREP messages through the source node. Eventually, \textit{a} starts to send data packets via the route drawn on the backbone.

In long-distance communication, since the routing process is started on-demand, routing errors only appear when a stored route in the distance vector is not valid anymore. In this case, similar to standard AODV algorithm, route error (RERR) messages are sent through the source node and the routing processes is retriggered. When the backbone is directly used for route discovery, the related route would be always discovered unless the cluster that destination node belongs is isolated.

\section{Results and Discussion} \label{sec:results}
All tests are conducted on discrete event simulator OMNeT++ and the simulation parameters in shown in Table \ref{tab:params}. To represent other cluster-based routing algorithms which focuses on the backbone to carry all packets, we implemented backbone routing upon clustered structure. We used standard AODV algorithm working in flat topology to emphasize advantages and disadvantages of clustering for routing protocols. For all three routing algorithms, a clustering algorithm which considers node degree, energy and connectivity to form clusters and decide roles (i.e CH, gateway and ordinary node) is used. While performance issues and internal details of the clustering algorithms is not in the scope of this study, the fairness of comparison between the routing algorithms is ensured using the exactly same clustering scheme for all of them.

\begin{table}[ht!]
		\centering
		\caption{The values of the simulation parameters.}
		\label{tab:params}
		\begin{tabular}{|l|l|l|}
			\hline
			\textbf{Parameter} & \textbf{Value}  \\ \hline
			Area size & $200\times 200$ m$^2$  \\ \hline
			Transmission area per node &  $\sim35$m  \\ \hline
			Node density &  $0.00125$ \\ \hline
			Mobility model &  Random Waypoint \\ \hline
			Ratio of mobile nodes & $30$\% \\ \hline
			Speed of nodes & $2-10$km/h \\ \hline
			Path loss model &  Free space path loss \\ \hline
			Power cons. model &  Radio state-based \\ \hline
			Background noise &  -90 dBm \\ \hline
			Runs per batch &  $200$ \\ \hline
			Scenario duration &  $60$s \\ \hline
		\end{tabular}
	\end{table}

In the study, ideal link layer and physical layer are considered to eliminate packet loss. To measure the success in data transfer, a UDP application which periodically (0.5 second) sends UDP packets between randomly selected source and destination nodes in uniformly distributed mobile topologies is implemented as well. 

There are five different performance metrics presented to discuss results. \textit{Average end-to-end delay} and \textit{packet delivery ratio between two random nodes} (PDR) are selected to examine performances of the different algorithms for data transfer. \textit{Standard deviation in power consumption} is important to reveal if only particular nodes such as members of the backbone are exhausted, or energy consumption is fairly distributed. To get more realistic results, radio state-based power consumption model is adjusted according to well-known chips Microchip RN1810 and Broadcom WSDB-102GN. \textit{Number of routing control packets} and \textit{size of routing control packets in bytes} show control overhead in terms of number of packets and total size in bytes. To measure the total size of control packets, different aspects and design issues are specifically considered for each algorithm. In the backbone routing, for the size of control packets flowing through the backbone, an optimum AODV packet size is selected, which is 64 bytes \cite{Hassan2011}. The same packet size is also chosen for the standard AODV algorithm. In contrast, even though the size of basic control packets is again 64 byte in CHRA (i.e route request, response and repair packets), the cost of topology discovery to form cluster sight area is added to the overall control overhead for a fair comparison. Note that, more detailed analytical approaches to measure control overhead are also studied before \cite{Xue2006}\cite{Abboud2011}, however since our simulation technique makes tracking and collecting general statistics of packet transmission much more easy in practical scenarios, we preferred to use such statistics to present actual overhead.

\begin{figure*}[htpb!]
\label{fig:uniform}
\centering
	\begin{subfigure}[t]{0.5\textwidth}
		\includegraphics[width=1\columnwidth]{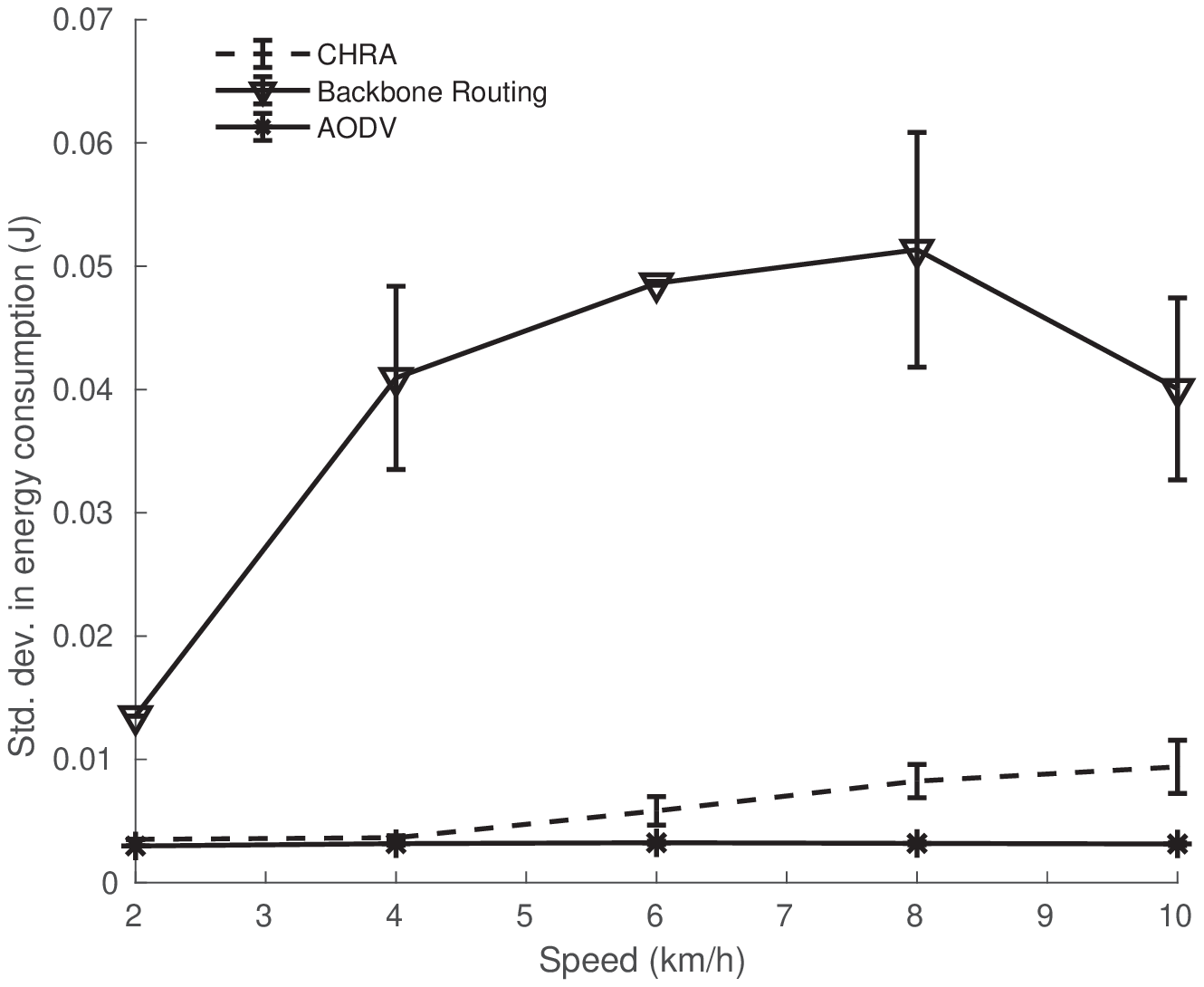}
		\caption{Standart deviation in energy consumption}
		\label{fig:aec_uni}
	\end{subfigure}%
	\begin{subfigure}[t]{0.5\textwidth}
		\includegraphics[width=1\columnwidth]{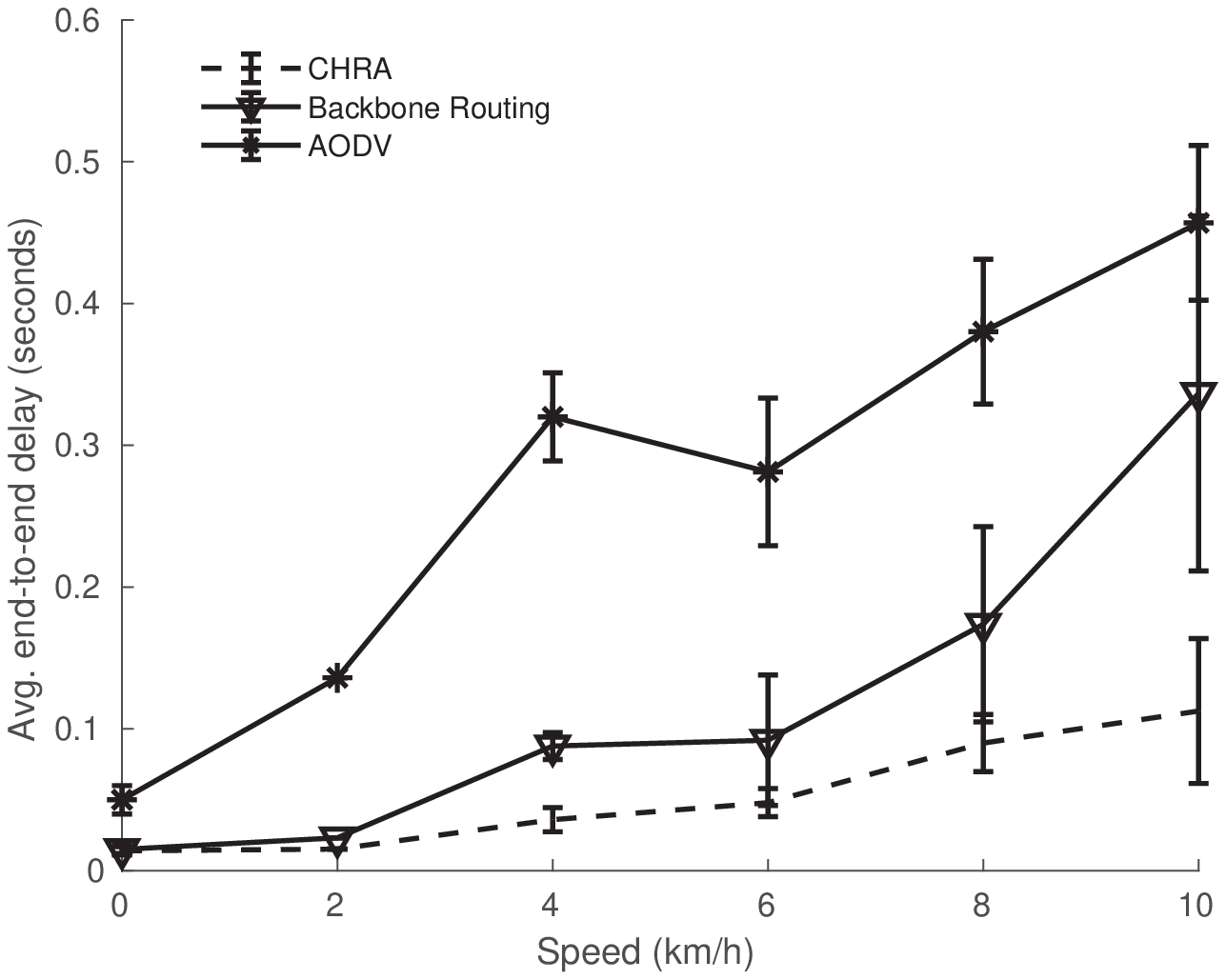}
		\caption{Average end-to-end delay (ms)}
		\label{fig:e2e_uni}
	\end{subfigure}%
	~ 	
	
	\begin{subfigure}[t]{0.5\textwidth}
		\includegraphics[width=\columnwidth]{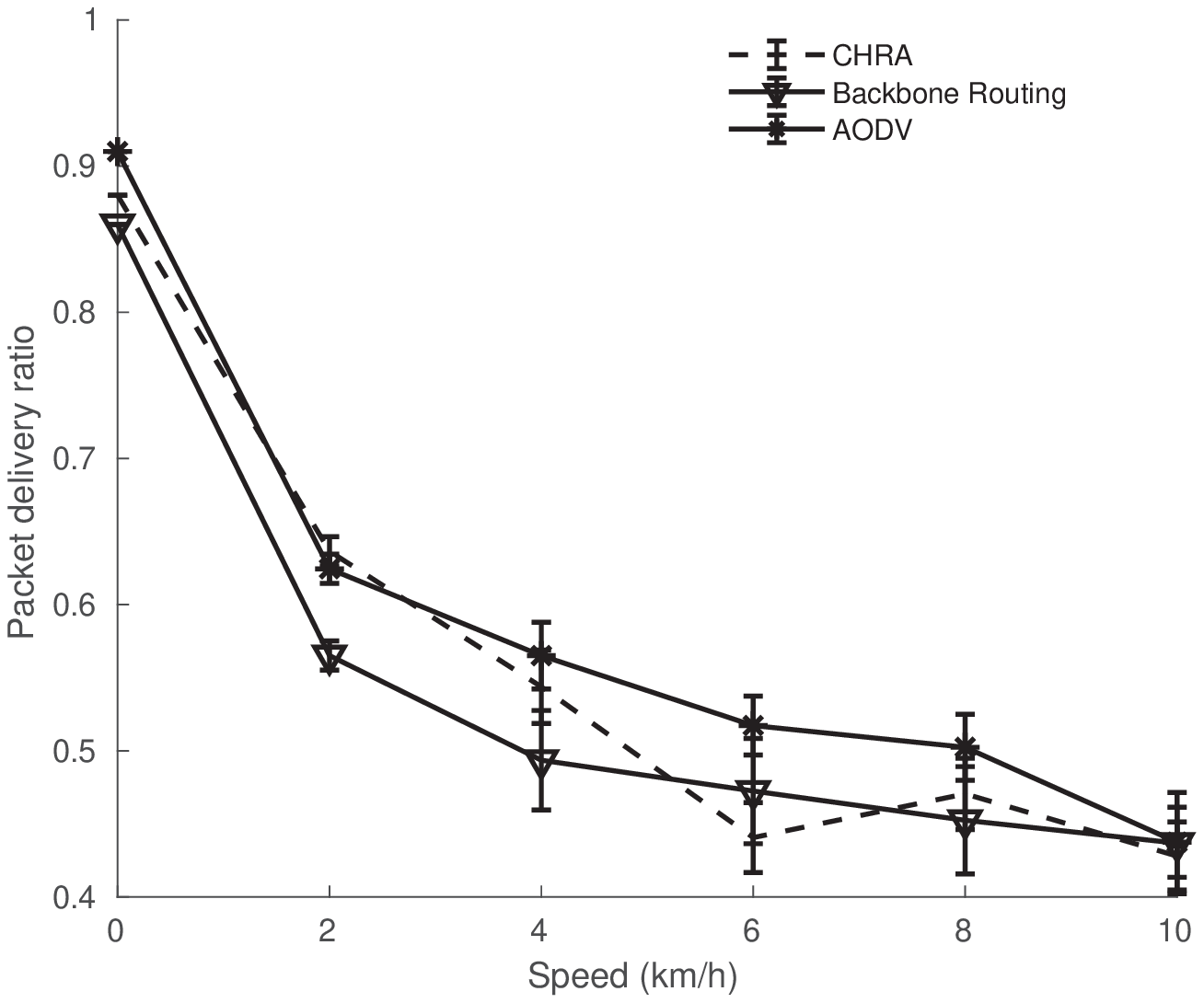}
		\caption{Packet delivery ratio}
		\label{fig:pdr_uni}
	\end{subfigure}%
	\begin{subfigure}[t]{0.5\textwidth}
		\includegraphics[width=\columnwidth]{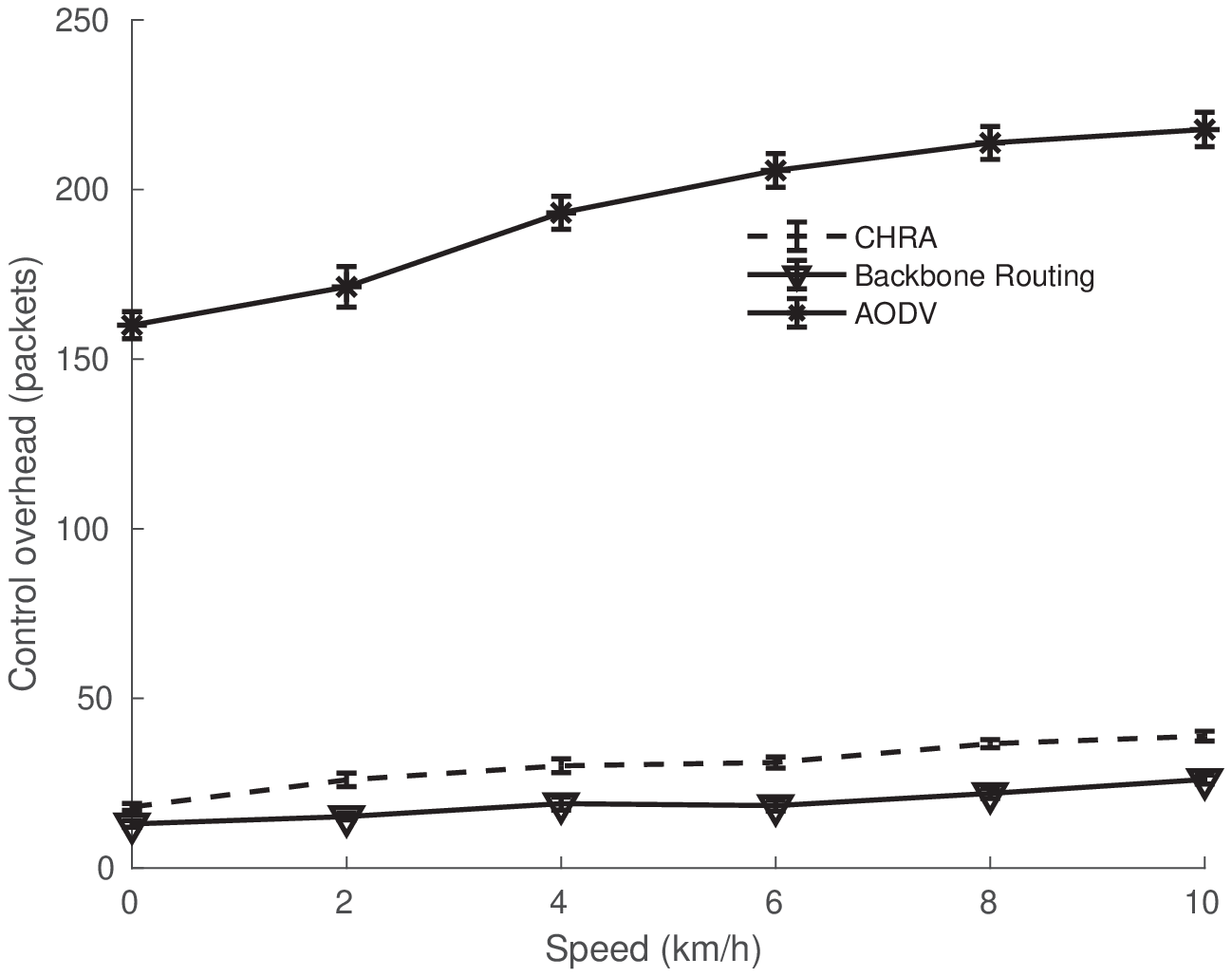}
		\caption{Number of control packets}
		\label{fig:ovr_uni}
	\end{subfigure}%
	~ 	
	
	\begin{subfigure}[t]{0.5\textwidth}
		\includegraphics[width=\columnwidth]{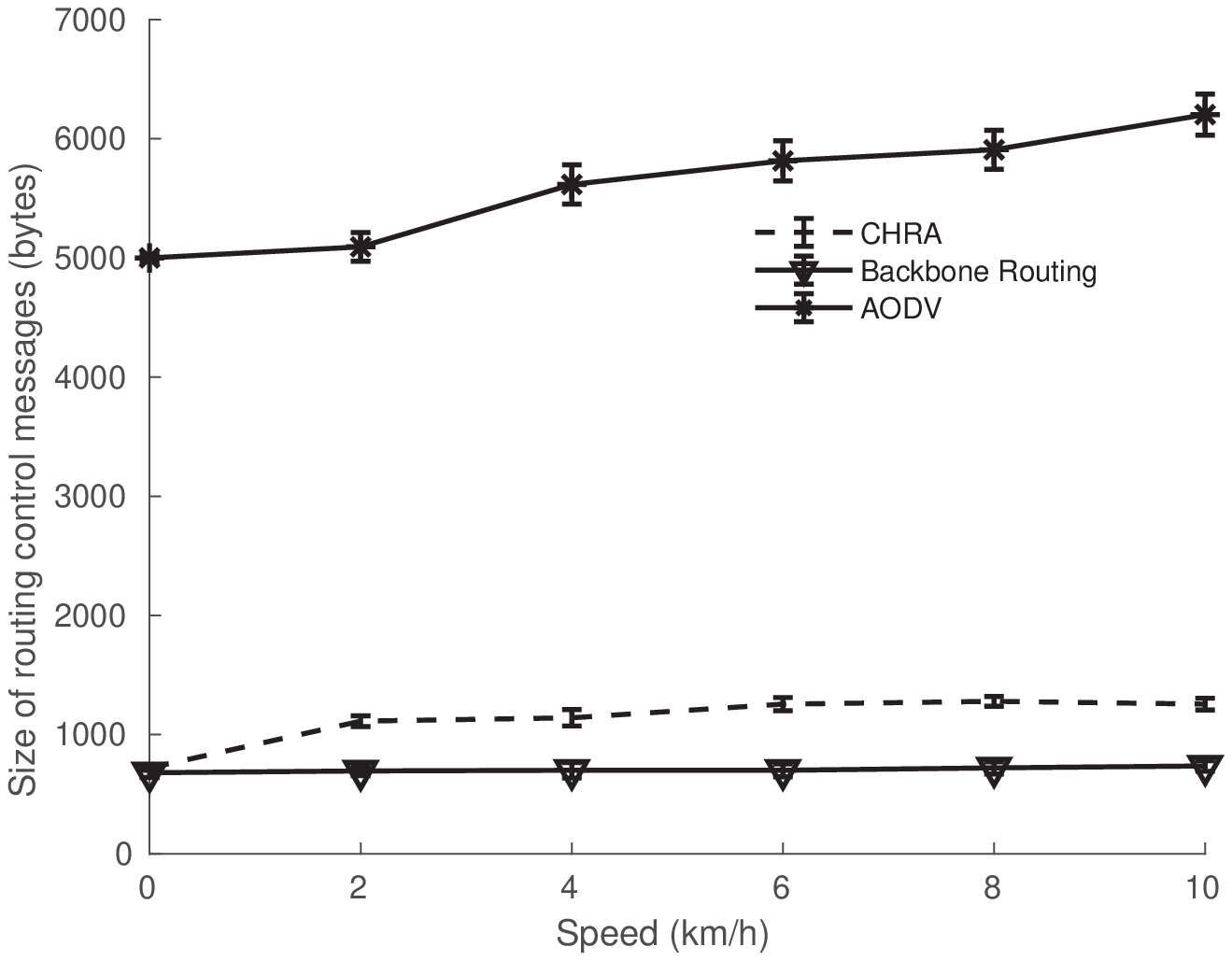}
		\caption{Size of control packets}
		\label{fig:size_uni}
	\end{subfigure}%
\caption{The effects of the speed in randomly distributed scenarios}
\end{figure*}

As shown in Fig. \ref{fig:pdr_uni}, PDR is very close to each other for all routing algorithms, and it varies between 40-90\% (including stationary scenarios where node speed is $0$km/h) depending on node speed. In contrast, there is an enormous gap in number of processed routing control packets. Fig. \ref{fig:ovr_uni} shows that, while AODV has a huge overhead due to broadcasting routing packets, cluster-based routing protocols has much less overhead using the backbone. The main reason for the difference between CHRA and backbone routing is, route error and repair techniques. However, we can observe the advantages of such repair technique in Fig. \ref{fig:e2e_uni}. In the figure, since CHRA is able to find alternative paths other than the backbone and repair invalid links quickly, its end-to-end delay performance is better than its opponents. Similarly, using the paths consist of alternative nodes rather than CHs and gateways, CHRA provides a fair energy consumption behavior between nodes. Fig. \ref{fig:aec_uni} reveals that, the nodes in backbone routing shows much more variation in energy consumption since only particular nodes (e.g CHs and gateways) are exhausted due to both control and data packets. In contrast, since the number of broadcast packets is significantly higher in AODV, nodes consume much more energy but in a fairer manner: they all consume high energy. 

Even if Fig. \ref{fig:size_uni} shows a quite parallel pattern with Fig. \ref{fig:ovr_uni} and, the difference between backbone routing and CHRA is higher in terms of total bytes for control messaging due to maintenance of CSAs and forwarding EEPs. Considering the gain in energy consumption and end-to-end delay, this extra overhead may be tolerable in many scenarios.

As Fig. \ref{fig:pdr_uni} indicates, PDR does not exceed 70\% in mobile scenarios; the reasons behind it are (1) occasionally isolated nodes due to mobility, (2) non-repairable routes in a short time due to mobility and (3) randomness in uniform distribution. We run the simulations in nonuniformly distributed mobile scenarios as well. We observed 80-50\% PDR in nonuniformly distributed mobile scenarios depending on the speed. Due to page-length limitations, we defer those results for an extended study.

	\begin{figure}[h]
	\centering
		\includegraphics[scale=0.7]{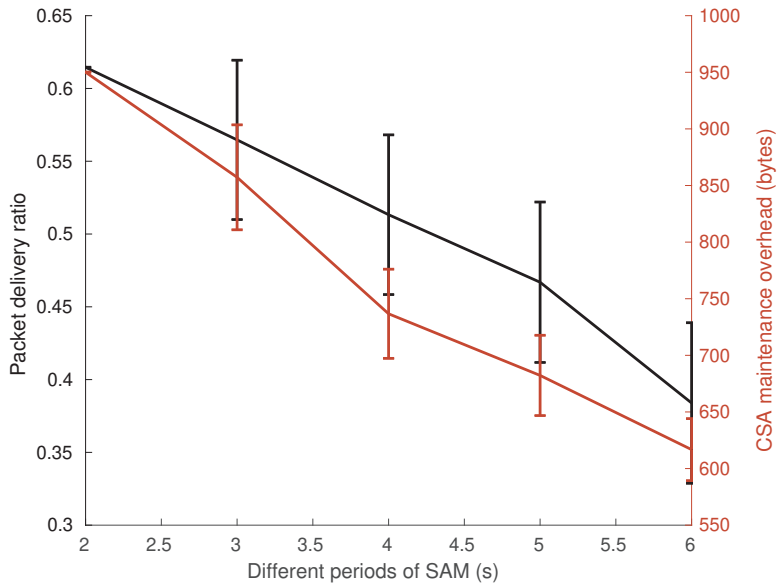}
		\caption{Effects of $T_{SAM}$ on packet delivery ratio and overhead}		
		\label{fig:tdiscuss}
	\end{figure}

Lastly, the control overhead for CSA maintenance is worth discussing. As seen in Fig. \ref{fig:tdiscuss}, PDR is decreasing with increasing period of SAM packets, $T_{SAM}$. In contrast, control overhead is getting less with more infrequent SAM packets as expected. The reason is, infrequent SAM packets directly lead to routing based-on obsolete topology information, and packets cannot be forwarded through destination when repair is not possible. In this manner, $T_{SAM}$ need to be decided based-on mobility characteristics of the network.

\section{Conclusion and Future Work} \label{sect:conclusion}
In this work, we presented a plane-separated hybrid routing algorithm in ad-hoc networks. The whole picture and the major dynamics of the algorithm in different scenarios are discussed. The separation of control plane and user plane leads to find alternative routes which is not dependent to the backbone in contrast to many other cluster-based routing algorithms. Using those alternatives provides a fair energy consumption scheme since a significant data forwarding burden is taken from control plane, and distributed to other nodes which triggers effective use of the user plane. The results also show that, with a proper repairing mechanism and using alternative paths in user plane, CHRA can handle data transfer with a lower delay while maintaining a high-level packet delivery performance. For future work, we are planning to discuss the quality of service in different network conditions considering packet loss, bit errors, interference etc., and also consider different traffic demands by a variety of applications. Lastly, comparing CHRA with additional routing algorithms in nonuniform distribution scenarios will be focused for extension of the study.

	\section*{Acknowledgment}
	This work is partially supported by ASELSAN.

\end{document}